\documentclass[twoside,twocolumn,9pt]{article}
\pdfoutput=1
\usepackage{extsizes}
\usepackage[super,sort&compress,comma]{natbib} 
\usepackage[version=3]{mhchem}
\usepackage[left=1.5cm, right=1.5cm, top=1.785cm, bottom=2.0cm]{geometry}
\usepackage{balance}
\usepackage{times,mathptmx}
\usepackage{sectsty}
\usepackage{graphicx} 
\usepackage{lastpage}
\usepackage[format=plain,justification=justified,singlelinecheck=false,font={stretch=1.125,small,sf},labelfont=bf,labelsep=space]{caption}
\usepackage{float}
\usepackage{fancyhdr}
\usepackage{fnpos}
\usepackage[english]{babel}
\addto{\captionsenglish}{%
  
}
\usepackage{array}
\usepackage{droidsans}
\usepackage{charter}
\usepackage[T1]{fontenc}
\usepackage[usenames,dvipsnames]{xcolor}
\usepackage{setspace}
\usepackage[compact]{titlesec}
\usepackage{hyperref}

\usepackage{epstopdf}

\usepackage{siunitx}
\usepackage{amssymb,amsmath}

\definecolor{cream}{RGB}{222,217,201}

\begin{document}

\pagestyle{fancy}
\thispagestyle{plain}
\fancypagestyle{plain}{

\renewcommand{\headrulewidth}{0pt}
}

\makeFNbottom
\makeatletter
\renewcommand\LARGE{\@setfontsize\LARGE{15pt}{17}}
\renewcommand\Large{\@setfontsize\Large{12pt}{14}}
\renewcommand\large{\@setfontsize\large{10pt}{12}}
\renewcommand\footnotesize{\@setfontsize\footnotesize{7pt}{10}}
\makeatother

\renewcommand{\thefootnote}{\fnsymbol{footnote}}
\renewcommand\footnoterule{\vspace*{1pt}%
\color{cream}\hrule width 3.5in height 0.4pt \color{black}\vspace*{5pt}} 
\setcounter{secnumdepth}{5}

\makeatletter 
\renewcommand\@biblabel[1]{#1}            
\renewcommand\@makefntext[1]%
{\noindent\makebox[0pt][r]{\@thefnmark\,}#1}
\makeatother 
\renewcommand{\figurename}{\small{Fig.}~}
\sectionfont{\sffamily\Large}
\subsectionfont{\normalsize}
\subsubsectionfont{\bf}
\setstretch{1.125} 
\setlength{\skip\footins}{0.8cm}
\setlength{\footnotesep}{0.25cm}
\setlength{\jot}{10pt}
\titlespacing*{\section}{0pt}{4pt}{4pt}
\titlespacing*{\subsection}{0pt}{15pt}{1pt}

\fancyfoot{}
\fancyfoot[LO,RE]{\vspace{-7.1pt}\includegraphics[height=9pt]{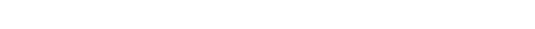}}
\fancyfoot[CO]{\vspace{-7.1pt}\hspace{13.2cm}\includegraphics{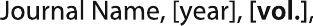}}
\fancyfoot[CE]{\vspace{-7.2pt}\hspace{-14.2cm}\includegraphics{RF}}
\fancyfoot[RO]{\footnotesize{\sffamily{1--\pageref{LastPage} ~\textbar  \hspace{2pt}\thepage}}}
\fancyfoot[LE]{\footnotesize{\sffamily{\thepage~\textbar\hspace{3.45cm} 1--\pageref{LastPage}}}}
\fancyhead{}
\renewcommand{\headrulewidth}{0pt} 
\renewcommand{\footrulewidth}{0pt}
\setlength{\arrayrulewidth}{1pt}
\setlength{\columnsep}{6.5mm}
\setlength\bibsep{1pt}

\makeatletter 
\newlength{\figrulesep} 
\setlength{\figrulesep}{0.5\textfloatsep} 

\newcommand{\topfigrule}{\vspace*{-1pt}%
\noindent{\color{cream}\rule[-\figrulesep]{\columnwidth}{1.5pt}} }

\newcommand{\botfigrule}{\vspace*{-2pt}%
\noindent{\color{cream}\rule[\figrulesep]{\columnwidth}{1.5pt}} }

\newcommand{\dblfigrule}{\vspace*{-1pt}%
\noindent{\color{cream}\rule[-\figrulesep]{\textwidth}{1.5pt}} }

\makeatother

\twocolumn[
  \begin{@twocolumnfalse}
{\includegraphics[height=30pt]{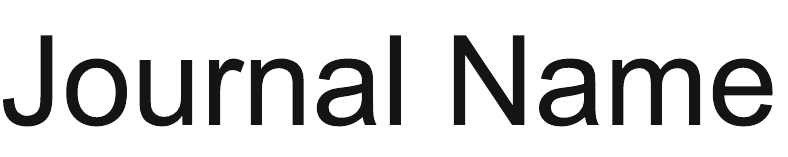}\hfill%
 \raisebox{0pt}[0pt][0pt]{\includegraphics[height=55pt]{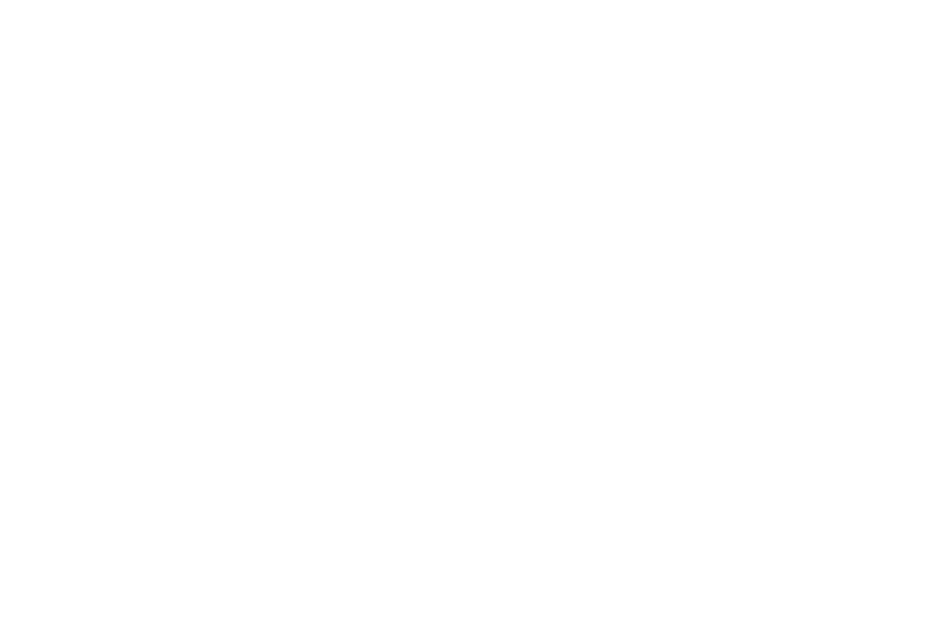}}%
 \\[1ex]%
 \includegraphics[width=18.5cm]{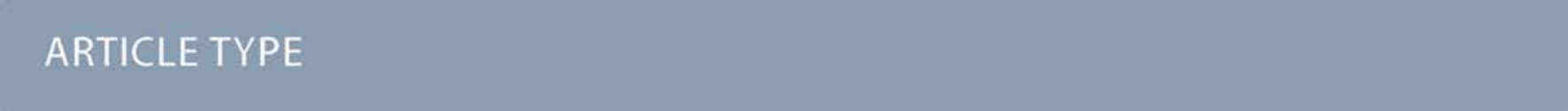}}\par
\vspace{1em}
\sffamily
\begin{tabular}{m{4.5cm} p{13.5cm} }

\includegraphics{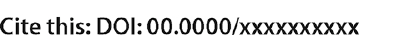} & \noindent\LARGE{\textbf{Soft matter science and the COVID-19 pandemic}} \\
\vspace{0.3cm} & \vspace{0.3cm} \\

 & \noindent\large{Wilson C K Poon,$^{\ast}$\textit{$^{a}$} Aidan T Brown,\textit{$^{a}$}  Susana O. L. Direito,\textit{$^{a}$} Daniel J M Hodgson,\textit{$^{a}$} Lucas Le Nagard,\textit{$^{a}$} Alex Lips,\textit{$^{a}$} Cait E MacPhee,\textit{$^{a}$} Davide Marenduzzo,\textit{$^{a}$}  John R Royer,\textit{$^{a}$} Andreia F Silva,\textit{$^{a}$} Job H J Thijssen,\textit{$^{a}$} and Simon Titmuss,\textit{$^{a}$} } \\

\includegraphics{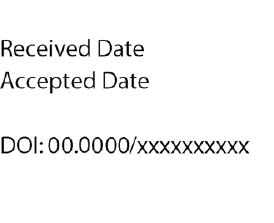} & \noindent\normalsize{Much of the science underpinning the global response to the COVID-19 pandemic lies in the soft matter domain. Coronaviruses are composite particles with a core of nucleic acids complexed to proteins surrounded by a protein-studded lipid bilayer shell. A dominant route for transmission is via air-borne aerosols and droplets. Viral interaction with polymeric body fluids, particularly mucus, and cell membranes control their infectivity, while their interaction with skin and artificial surfaces underpins cleaning and disinfection and the efficacy of masks and other personal protective equipment. The global response to COVID-19 has highlighted gaps in the soft matter knowledge base. We survey these gaps, especially as pertaining to the transmission of the disease, and suggest questions that can (and need to) be tackled, both in response to COVID-19 and to better prepare for future viral pandemics.} \\

\end{tabular}

 \end{@twocolumnfalse} \vspace{0.6cm}

  ]

\renewcommand*\rmdefault{bch}\normalfont\upshape
\rmfamily
\section*{}
\vspace{-1cm}


\footnotetext{\textit{$^{a}$~Edinburgh Complex Fluids Partnership (ECFP), SUPA and School of Physics \& Astronomy, The University of Edinburgh, Peter Guthrie Tait Road, Edinburgh EH9 3FD, United Kingdom. E-mail: w.poon@ed.ac.uk}}





The coronavirus disease 2019 (COVID-19) pandemic caused by the severe acute respiratory syndrome coronavirus 2 (SARS-CoV-2)\cite{Rabadan2020} has focussed unprecedented attention on science and technology. Fighting a pandemic is, at first sight, a challenge principally for biotechnology and the biomedical sciences, which have indeed responded rapidly: witness, for example, the speed at which candidate vaccines have been brought to clinical trial. However, the reality is that effort from many disciplines is needed to respond adequately to the pandemic. Thus, the sudden need for extra ventilators has brought innovative solutions from engineering design and manufacturing. Ventilation is only one aspect of the fluid dynamics needed to confront COVID-19. Other fluid dynamical aspects of the disease, such as the aerodynamics of aerosol transport, have been reviewed.\cite{Mittal2020} 

Coronaviruses belong to the family of `enveloped viruses'. An enveloped virus has a lipid bilayer `shell' with embedded proteins enclosing a `core' consisting of nucleic acids complexed with proteins.\cite{Strauss2008} It is therefore a `quintessential  soft matter object': a composite colloid made up of surfactants and polymers. Respiratory coronaviruses such as SARS-CoV-2 are transmitted\cite{Zhang2020} by another kind of soft matter object, aerosols in which the liquid phase is rich in mucin and other biopolymers. Here, we survey some of the soft matter science that is relevant to COVID-19, paying particular attention to gaps in the knowledge base. In doing so, we also want to provide entry points into a diverse literature, but with no claim to completeness. The review already cited\cite{Mittal2020}, a critical compilation of numerical data,\cite{Milo2020} an overview of the challenges presented to physical scientists and engineers by COVID-19,\cite{Huang20} and a survey of nanotechnological responses to the pandemic\cite{Weiss2020} are also useful sources of references, as is a very recent paper emphasizing the multi-scale nature of the response needed.\cite{Bellomo2020} A web site updated every 24 hours provides an efficient way to keep up to date with a burgeoning literature,\cite{primer} while an interdisciplinary text provides a useful panoramic introduction to viral biophysics.\cite{mateu2013}

The soft matter science of SARS-CoV-2 falls naturally under two headings: how the virus invades the body, and how infection is spread. Both stories start with a virus on the surface of the respiratory tract, Fig.~\ref{fig:transmit}(a). The epithelia of our respiratory,\cite{Thornton2008}  gastrointestinal and reproductive tracts are covered by viscoelastic mucus. Its composition varies with site, time and state of health, but the most important macromolecular components are high molecular weight mucin proteins and DNA shed from cell debris.\cite{Mackie2014,Bansil2018} 

\section{The `inside' story}

The `inside' story of viral transmission starts with a virus landing on a mucosal surface.\cite{Sato2012} The system of beating cilia on mucosa may clear viruses away.\cite{Zanin2016} The emergent field of `active matter' has contributed much to understanding the coarse-grained physics of cilia dynamics, e.g., the role of hydrodynamic interactions in the generation of collective beating.\cite{Gilpin2020} However, the study of the physics of `mucociliary clearance'\cite{Marin2017}  -- how propagating `metachronal waves' transport mucus and convey trapped pathogens out of the body -- is only just beginning.\cite{Chateau2019}

Viruses not expelled by the mucociliary clearance apparatus then have to diffuse through\cite{Olmsted2001,Lai2010,Wang2017,Metzler2019} a highly heterogeneous viscoelastic porous medium.\cite{Mackie2014,Lai2010} This involves the generic physics of nanoparticle diffusion in soft porous media\cite{Borros2015}  and the more specific physics of adhesive receptor-ligand binding, both of which are also relevant for designing synthetic nanoparticles for drug delivery.\cite{Liu2015}  As far as adhesion is concerned, bacteria behave as colloids with sticky patches, and show significant phenotypic heterogeneity.\cite{Vissers2018} On statistical grounds, it seems likely that the distribution of sticky moieties is also patchy on enveloped viruses, so that considering them as patchy nanoparticles\cite{Banquy2019} may provide new insights, e.g. concerning wettability,\cite{Velev2016} which may influence mucosal penetration. Virions that succeed in diffusing through the mucus then face an osmotic permeability barrier due to mucins tethered to epithelial cells.\cite{Button12} 

Beyond this first stage of mucosal penetration, the `inside' story rapidly becomes dominated by specific virus-cell interactions. The literature here is itself dominated by a fine-grained approach focussing on molecular details. However, a more coarse-grained, soft matter approach can also make important contributions. One particularly important area for our purposes in which this is undoubtedly true is the process in which enveloped viruses gain entry to host cells. (See Chapter 16 in Mateu's text\cite{mateu2013} by M\'as and Melero for a general introduction, and Hoffmann et al.\cite{Hoffmann2020} for SARS-CoV-2.) For example, the proteins embedded in the outer lipid bilayer envelope play a crucial part in altering membrane curvature to facilitate fusion.\cite{Basso2016} Moieties that can change the local curvature therefore have potential to be antiviral drugs.\cite{StVincent} Soft-matter-inspired approaches that neglect much of the atomic details can also contribute to the identification of potential binding sites in viral proteins. Thus, e.g., a recent computational study of the dynamics and thermodynamics of the main protease of SARS-CoV-2 from the perspective of the biophysics of fluctuation elasticity in globular proteins\cite{McLeish2020} has identified new candidate sites for the binding of inhibitors, with clear pharmacological implications.
 
\begin{figure*}[t]
\centering
  \includegraphics[height=11cm]{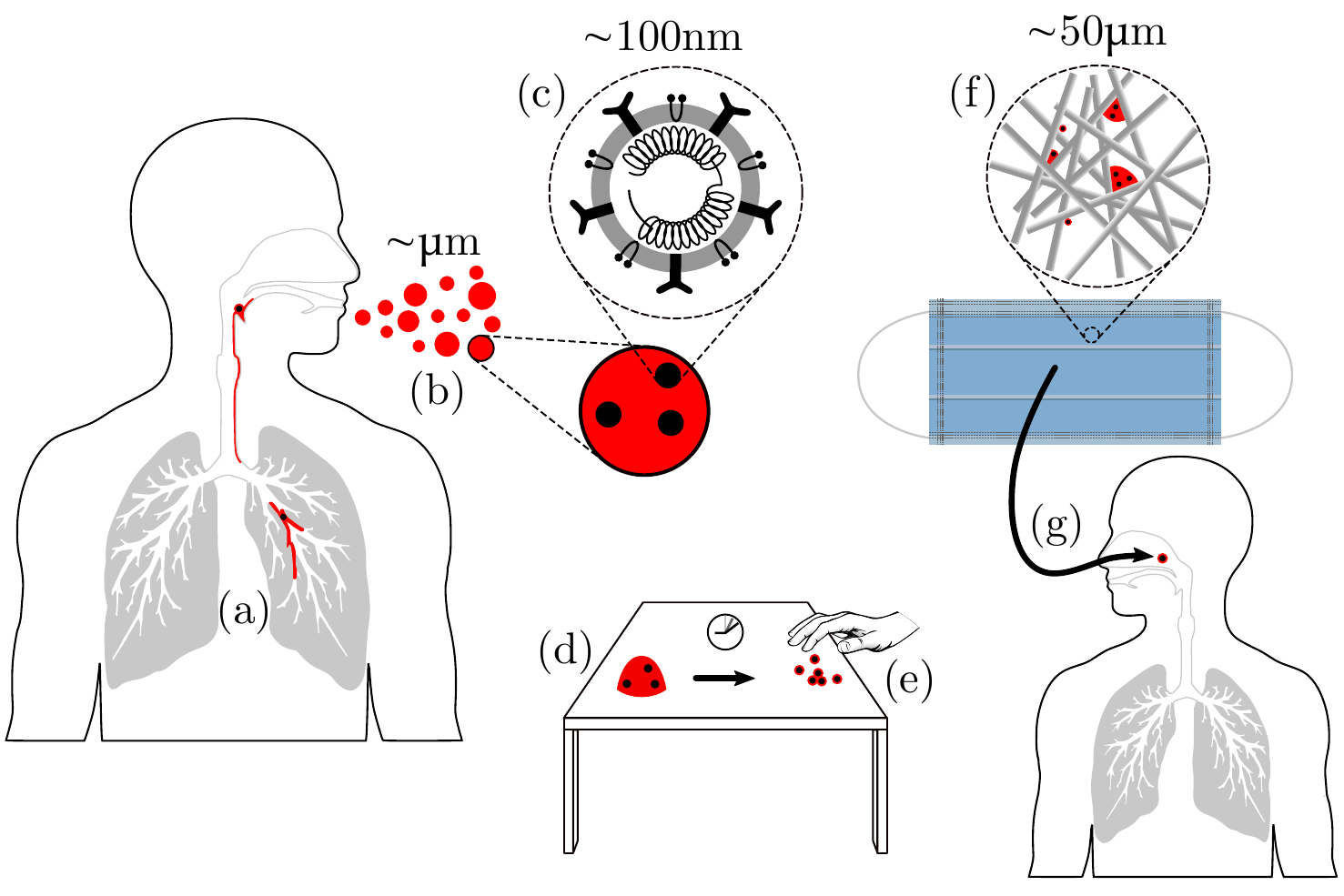}
  \caption{Schematic of some of the soft matter science in virus transmission (not to scale!). (a) Virus particles (black) are presented on or in the mucus lining of internal epithelia (red). (b) Droplets of virus-bearing mucus are ejected via coughing, sneezing, breathing or talking. (c) A coronavirus with RNA surrounded by a lipid bilayer in which are embedded various proteins. (d) Droplets landing on surfaces will dry while exposed to air. (e) The final result is adsorbed viral particles, individual or clustered, each presumably with residual mucosal biopolymers, lipids and salts. Hands touching such surfaces may pick up virions and spread infection. (f) A droplet impinges onto a face mask, whose microstructure is a network of polymer fibres; these trap droplets, which then dry to leave adsorbed viruses, again complexed with biopolymers, lipids and salts. (g) Air flow (breathing) or liquid flow (e.g.~washing) through this complex porous medium determines its effectiveness and susceptibility to cleaning. }
  \label{fig:transmit}
\end{figure*}

Much more can be said about the soft matter science of the `inside story'; but that will require a full essay of its own. For the rest of this essay, we will focus instead on the `outside' story of SARS-CoV-2, where biologically specific interactions play less of a role (but see a recent paper on multi-scale modelling\cite{Bellomo2020} for additional references on the `inside' story). We will first discuss the soft matter science of SARS-CoV-2 transmission and of protecting ourselves against the virus, Fig.~\ref{fig:transmit}. Two briefer sections then survey how the virus or infection can be detected and the sustainability issues raised by COVID-19.  We end with a general discussion and some concluding remarks.  

\section{Transmission}

Airborne transmission is the dominant route for the spread of COVID-19.\cite{Zhang2020} A simplified schematic of some of the components of airborne transmission is shown in Fig.~\ref{fig:transmit}. What is immediately obvious is that a challenging range of length scales is involved.\cite{Bellomo2020} Soft matter scientists deal regularly with such multi-scale problems, because soft matter inhabits the `middle world' that bridges the microscopic and the macroscopic.\cite{Haw2016} In terms of the average diameter of a coronavirus\footnote{The size and shape of viruses is phenotypically heterogeneous, especially among enveloped viruses; such `pleomorphism' may be an adaptive trait.\cite{Li2020}} $D \approx \SI{100}{\nano\meter}$, the relevant length scales span from $\sim 10^{-2}D$ (individual coat proteins) through $10^2D$ (fabric microstructure) to $10^7D$ (macroscopic air flow). More implicit, but no less challenging, is the range of relevant time scales, spanning from the Brownian time of a single virus in water (time taken to diffuse its own diameter), $\tau_{\rm B} \lesssim \SI{1}{\milli\second}$, through the $\sim 10^4\tau_{\rm B}$ needed for \SI{0.1}{\milli\metre} droplets to evaporate completely in air\cite{Wells1934,Xie2007} and the $\lesssim 10^5\tau_{\rm B}$ for virus-cell fusion processes,\cite{Floyd2008} to the $\gtrsim  10^9\tau_{\rm B}$ (weeks) taken by COVID-19 to run its course in humans.\cite{Wolfel2020}  

\subsection{Air-borne droplets} \label{sec:air}

Respiratory viruses are transmitted from human to human either via the air or via contact, the latter either directly via an infected individual or indirectly via a contaminated surface.\cite{Kutter2018} Virus-bearing mucus is brought up and expelled as smaller `aerosols' or larger `droplets' when an individual exhales, speaks, coughs or sneezes, Fig.~\ref{fig:transmit}(a,b). The distinction between aerosols and droplets\footnote{In some aerosol science literature, what we call `aerosols' here are called `droplet nuclei', and both droplets and droplet nuclei are known as aerosols.\cite{Hsiao2020}} originated with W. F. Wells' 1934 work that studied the fate of water drops in air. He proposed a crossover  diameter below which a drop would completely evaporate before it fell to the ground, while a bigger drop would hit the ground before it had completely evaporated.\cite{Wells1934} This crossover clearly depends on the initial height at which the drop is released, its initial motion, and the relative humidity of the air. The original estimate\cite{Wells1934} of $\lesssim \SI{200}{\micro\meter}$ has been revised down,\cite{Xie2007} and in the current virology literature is often taken to be $\approx \SI{5}{\micro\meter}$ (see, e.g., Zhou et al.\cite{Zhou2018}). We will see shortly that this `one-body' picture does not capture the complexity of the real situation. We will therefore not make the aerosol/droplet distinction, and use the word `droplet' as a generic term covering all sizes, mentioning the diameter explicitly if it is relevant. Using the term in this sense, we learn from a study of the droplets produced by coughing\cite{Yang2007} that the size distribution spans a wide range, from $\approx 0.6$ to $\approx \SI{16}{\micro\meter}$, with a mode of around \SI{6}{\micro\meter}. 

Respiratory droplets expelled by infectious individuals can be directly inhaled by another person or deposited on surfaces, either on another person or on environmental objects (which, insofar as they carry infection, are known as `fomites'); anyone touching fomites and then the mucous membranes of their own eyes, noses or mouths may become infected. 

The droplets ejected by sneezing, etc.\cite{Bourouiba2014,Scharfman2016,Mittal2020} contain mature viral particles, or virions, dispersed in a solution of inorganic salts (NaCl, etc.), surfactants (lipids), mucins, and probably other biopolymers,\cite{Marr2018} Fig.~\ref{fig:transmit}(c). This compositional complexity is important. For example, the presence of mucin is known to improve the survivability of the H1N1 flu virus in aerosols and droplets,\cite{Otter2016} rendering it more or less independent of the relative humidity of the environment.\cite{Marr2018a} To underline this compositional complexity, we will speak of biopolymer-lipid-salt-virion (BLSV, pronounced  `BiLi-SaVi') droplets and deposits on surfaces.

Intuitively, one may imagine that virus bearing material is ejected by infected individuals as already-formed droplets -- indeed, we have seen that this `one-body' picture lies behind the distinction between `aerosols' and `droplets'. However, recent fluid dynamical work\cite{Mittal2020,Bourouiba2014,Scharfman2016} shows that at least in sneezing, respiratory droplets are not primarily ejected `as formed', but are the result of multiple fragmentation processes post-ejection. It is known that in Newtonian fluids, particles, even when present at low dilution, can decisively influence the kinetics of jet and sheet fragmentation (see Lindner et al.\cite{Lindner2015} and references therein for jets and Raux et al.\cite{Raux2020} for sheets), although the distribution of particles in the resulting population of droplets has not, to our knowledge, been studied. Moreover, threadlike structures observed in sneeze ejecta\cite{Scharfman2016} implicate high-molecular-weight mucins, because such threads are characteristic of fluids with significant elasticity.\cite{Bhat2010} Importantly, it is not so much the `single-body' motion of isolated droplets but  the collective motion of a propelled turbulent cloud of droplets that controls subsequent deposition on surfaces.\cite{Mittal2020,Bourouiba2014,Scharfman2016} 

A recent article entitled `COVID-19 by the numbers'\cite{Milo2020} has highlighted the importance of quantification. A good first question for a quantitative soft-matter approach to SARS-CoV-2 transmission is: how many viruses are there per droplet? A study of the distribution of virus numbers in aerosols artificially generated by nebulisation\cite{Zuo2013} finds concentrations that extrapolate to about one virion per $\sim \SI{1}{\micro\meter}$ droplet.\footnote{However, this and similar studies involve nebulising virions using Newtonian aqueous solutions and therefore do not  reproduce the mucoidal ejecta from which real-life droplets are generated.} Other studies have found that the majority of droplets carrying influenza viruses may be $\lesssim \SI{5}{\micro\meter}$ in diameter (see Bischoff et al.\cite{Bischoff2013} and references therein), while a study of SARS-CoV-2 in hospital environments\cite{Liu2020} reports a more variable picture, even though in some cases, the viral load is still concentrated in small (in this case $\lesssim \SI{1}{\micro\meter}$) droplets. 

However, there appears to be no measurement of viral titre in individual human respiratory droplets. Nevertheless, the figure of `100,000 to 1,000,000 virions per droplet' for the flu virus is widely quoted.\cite{Shors2017} As we will see shortly, this figure is wildly unlikely. It possibly originated as an estimated upper bound based on maximum packing:\cite{Weber2008} a \SI{10}{\micro\meter} droplet can contain up to $\sim~(10/0.1)^3 = 10^6$ particles of $\SI{0.1}{\micro\meter}$ diameter, corresponding to a viral volume fraction of $\phi_{\rm v} \sim 1$. 

For an evidence-based estimate, we turn to a recent measurement of the concentration of the viral load in the sputum of hospitalised patients with COVID-19.\cite{Wolfel2020} This study found that the average viral titre was $\approx \SI{7e6}{\per\milli\litre}$, with a maximum of $\approx \SI{2e9}{\per\milli\litre}$. Another recent study of the saliva of \mbox{COVID-19} patients found loads of $\lesssim \SI{e6}{\per\milli\litre}$ within a week of onset of symptoms, and a peak load of $\lesssim \SI{e9}{\per\milli\litre}$ in a number of patients in their early 60s.\cite{To2020} Using a virion diameter of \SI{100}{\nano\meter}, the maximum load from the sputum study translates to $\phi_{\rm v} \approx 10^{-6}$. If such sputum is entirely turned into \SI{5}{\micro\meter} droplets and the viral load is distributed uniformly, then we expect the mean number of virions per droplet to be $\sim 0.1$. If a Poisson distribution applies, then the number of \SI{5}{\micro\meter} droplets with 0, 1, and 2 virions will be in the ratio 200:20:1, so that $\approx 1\%$ of \SI{5}{\micro\meter} droplets will contain (single) virions, and only 0.55\% of the droplets will have $\geq 3$ virions.\footnote{This calculation assumes that the distribution of particles in droplets is simply that obtained from the average particle concentration in the starting mucosal material. This assumption may not be valid because particles affect ejecta fragmentation, so that droplet size and particle concentration may be coupled. }  

A method for measuring the viral titre in respiratory droplets {\it in situ} will clearly enable more precision in this matter. Meanwhile, we should put our estimates in the context of two other statistics. First, a cough or talking for 5 minutes can generate $\sim 3 \times 10^3$ droplets,\footnote{However, note that there exists a subgroup in the population who are `speech superemitters' who can emit an order or magnitude more droplets than their peers.\cite{Asadi2019}.} while a sneeze can generate $\lesssim 4 \times 10^4$ (see the review by Cole and Cook\cite{Cole1998} and references therein). Secondly, the (strain-dependent) minimum infective dose (MID)\footnote{The minimum dose needed to cause infection in 50\% of individuals.} of the flu virus is typically a few thousand virions for influenza.\cite{Nikitin2014} 

The next soft matter question is: where are the virions in the respiratory droplets? The answer is poorly known at present.\cite{Marr2018} A \SI{100}{\nano\meter} virus takes only $\approx \SI{1}{\second}$ to diffuse from the centre of a \SI{10}{\micro\meter} water droplet to its surface. A virus or indeed any other particle approaching an air-water interface may become bound to the interface, with or without breaching it. The outcome depends on the salt concentration and the details of the particle surface.\cite{Vinny2019,Stocco2020} The observation of non-monotonic dependence of both the magnitude and sign of the particle-interface interaction on salt concentration\cite{Vinny2019} is particularly interesting, because this variable continuously increases as the droplet evaporates in transit and on surfaces. If virions do breach the air-water interface, then they may be subjected to strong interfacial forces (see Section~\ref{sec:interfacial}). 

The final cluster of questions is about evaporation. BLSV droplets start to evaporate immediately upon release into the air. A study using environmental chambers finds that the viability of bacteriophages suspended in droplets of growth media (salts + small molecule metabolites) shows a non-monotonic dependence on the relative humidity, displaying a pronounced minimum in viability at intermediate relative humidity.\cite{Lin2020} The authors suggest that this is because viral survival depends on the product of solute concentration and time. In other words, the evaporation {\it kinetics} of droplets matters. 

The evaporation kinetics of droplets of \SI{300}{\nano\meter} silica particles in salt solution deposited on porous superhydrophobic surfaces (to mimic airborne droplet conditions\cite{Liu2019}) was found to depend strongly on both salt and colloid concentration.\cite{Velev2014} The latter is likely very low in respiratory droplets according to our previous discussion of viral titre, so it is the biopolymer-lipid-salt solutes that will determine the evaporative kinetics of such droplets. A recent study of model BLSV  droplets suggests that concentration and pH changes during in-transit evaporation may lead to a core-shell structure and/or induce mucin gelation.\cite{Marr2018} The development of such structure will undoubtedly affect drying kinetics, as well as how such droplets impact environmental surfaces (see next section). It will also control, ultimately, the structure of the encrusted virions -- composite BLSV particles -- left behind on fomites, which, in turn, may affect viral survival. 

\subsection{Droplet-surface interaction} \label{sec:surfaces}

After expulsion from the body, BLSV droplets eventually impact a variety of surfaces, which become potential sites for transmitting infection,\cite{Kutter2018} Fig.~\ref{fig:transmit}(e). The relevant questions for soft matter science in fomite transmission can be discussed under three headings: impact, drying and removal. First, there is a need to understand the initial impact of the droplets on surfaces. Secondly, after impact, the droplets dry to leave a deposit of virions; so the kinetics of the drying process and the structure of the composite BLSV particles left behind need to be elucidated and their relevance for viral viability explored. Thirdly, these BLSV particles can be removed by cleaning or picked up on skin. The underlying mechanisms of these processes need to be understood. 

The impact of Newtonian liquid droplets on hard surfaces has been well studied.\cite{Thoroddsen2016} The effect of viscoelasticity in the droplet has attracted recent attention, partly because of the `anti-rebound' effect of high-molecular-weight polymeric additives.\cite{Bertola2013,Muradoglu2016} It is clearly of interest to know if the mucus in respiratory droplets bearing viruses also confers this property, and the possible effects of mucin gelation. The possibility of structuring at the air-liquid interface\cite{Marr2018} further complicates the likely behaviour at impact.

Post impact, droplets dry to leave BLSV particles on the surface, Fig.~\ref{fig:transmit}(e). The drying of a dilute droplet of spherical colloids in a Newtonian fluid on a featureless hard substrate is well understood: it gives rise to a `coffee ring' in which all the particles are deposited at the rim.\cite{Larson2002} Gelation of the evaporating droplet\cite{Haw2002} or surface-driven flows induced by the presence of surfactants\cite{Stone2016} can inhibit coffee ring formation and lead to a more uniform deposit of particles. Mucin gelation and/or the presence of lipids in respiratory viral droplets \cite{Marr2018} may have this effect. However, the small size of virions may mean that they can diffuse in a mucin gel network, and nullify the expected effect of gelation. Respiratory droplets also contain salts. The drying of salt solution droplets differs significantly from the drying of colloidal suspensions.\cite{Shahidzadeh2015} How this is modified by the presence of lipids and biopolymers is unknown. Note that there is a growing literature on the drying of blood droplets aimed at forensic science and personal health applications.\cite{Brutin2011,Chen2016} Since blood is a complex fluid of biological origin, this literature may give additional insight into the drying of BLSV droplets (e.g. the role of proteins). 

The environmental surfaces that mucoidal viral droplets may land on range from relative simple -- a glass table top -- to very complex, such as a fabric mask surface, Fig.~\ref{fig:transmit}(f), or human skin, where a network of microchannels controls surface fluid transport.\cite{Lips2003} The study of drying on such patterned and/or `soft' surfaces is an active area of soft matter research.\cite{Gerber2019} Given the cocktail of solutes, the final deposit on these simple or complex surfaces after drying is complete will not be individual bare viruses, but composite BLSV particles. The drying kinetics will determine the structure of these composites.  

Unsurprisingly, there exist many measurements of viral viability on fomite surfaces\cite{Kampf2020,Ren2020} If we look behind reported survival times to the raw data,\cite{Lai2005,Munster2020} we repeatedly see that, once deposited, the number of viable viruses decreases according to $n\left(t\right)=n_0\exp{\left[-t/t_0\right]}$, but with wide variations in the actual value of $t_0$ for different virus-surface combinations. Currently, this time scale is simply accepted as a `brute fact' to be determined experimentally. So, for example, for SARS-CoV-2, $t_0 \approx 1$ to \SI{2}{\hour} on copper and $\approx 7$ to \SI{8}{\hour} on plastics.\cite{Munster2020}  However, the fact that there seems always to be a well-defined, characteristic time scale irrespective of the specific virus and surface is immediately striking for the physicist, suggesting an (unknown) generic underlying mechanism.

Surface viability studies typically do not specify explicitly what kind of droplets are being deposited. For example, in one description of experimental method, we read that viral aerosols `were generated by passing air at a flow rate of 7.5 L/min through a 3-jet Collison nebulizer'.\cite{Fischer2016} One can only assume that this generated droplets of viruses in aqueous (probably isotonic saline) solutions. It is not obvious that such studies should be relevant to the survival of viruses encrusted in biopolymers, lipids and salts, either as individuals or in clusters. The presence of these moieties will affect viral survival for a number of reasons. 
	
First, ions are integrated into the structure of viral capsids, and it is known that  such `structural ions' are important for preventing capsid collapse during the desiccation of the Triatoma virus (TrV).\cite{Pedro2018} Presumably the same applies more generally to other viruses. Secondly, we may expect that biopolymers, lipids and salts could help retain residual moisture and therefore change the local absolute humidity (mass of water per unit volume of air). Both the absolute\cite{Shaman2009} and relative\cite{Kampf2020,Ren2020} humidity are known to affect viral survival. Much more work is clearly needed, e.g., to clarify the putative role\cite{Minhaz2010,hanley2010,Marr2019} of osmotic effects (see section~\ref{sec:osmo} below) in the observed dependence of viability on humidity.

To complete the transmission cycle, we need, finally, to consider surface-to-surface transfer.\cite{Otter2016} Many recent studies have measured the transfer of a variety of bacteria and viruses between fomites and hands or cleaning cloths (see Zhao et al.~\cite{Zhao19} and references therein). Of the wide range of environmental factors studied, a few have very strong effects, e.g., for some bacteriophages, humidity increases the rate of surface-to-hand transfer approximately threefold,~\cite{Lopez13} perhaps pointing to a role for capillarity. Surface roughness and porosity, the contact force, the direction of transfer (hand to surface or vice versa), and the type of microorganism are all also significant.\cite{Zhao19} 

Studies to date of the transference of particulate matter between surfaces have been largely empirical, producing parameters such as the transference efficiency between different types of surfaces\cite{Rhodes2001,Byrne2012,Zhao2018} as inputs for numerical modelling of the transference process.\cite{Zhao2018} There are therefore significant opportunities for soft matter science to contribute towards determining the mechanistic basis of surface to surface transfer. The study of single particle detachment from solid\cite{Sharma2001} or liquid surfaces\cite{Butt1994} using AFM is one possible avenue; probing the role of tackiness\cite{Leibler1999} in the transference of respiratory droplets is another. In this context, note that some of the relevant soft matter science should also be applicable to the transference of trace evidence in forensic science,\cite{Forensic2009} so that fundamental advances here will benefit multiple fields. 

\subsection{Forces} \label{sec:forces}

During transmission, virions are subjected to forces of many kinds. We now consider the origins and measurement of these forces, which in the right circumstances may lead to viral inactivation.

\subsubsection{Electrical}

Both nucleic acids and proteins are charged, so that electrostatics is important in viral physics.\cite{Zandi2020} For example, it needs to be accounted for to understand genome packaging,\cite{Podgornik2012} and controls the osmotic pressure differential between the virion interior and the external medium.\cite{Garmann10472} A simple estimate of the electrostatic self-energy of a viral capsid, viewed as a protein shell of uniform charge density (which can be as large as one electron per nm$^2$), shows that this scales as $\sigma^2(\epsilon_0\epsilon)^{-1}R^2\lambda$;\cite{Podgornik2012}  here $R$ and $\sigma$ are the capsid radius and surface charge density respectively, $\lambda$ is the Debye length, and $\epsilon_0$ and $\epsilon$ are the permittivity of vacuum and the dielectric constant of the medium. Using realistic numbers for RNA virions, we find self-energies of $\sim 10^4 k_{\rm B}T$ in physiological buffers (i.e., 150 mM monovalent salts such as NaCl), but the exact number is sensitive, for instance, to the salt concentration in the local viral environment. As virions self assemble, this self-energy needs to be balanced by hydrophobic or van der Waals interactions, and varying salt concentration may be a way to tilt the balance towards assembly or disassembly.\cite{Podgornik2012}  These considerations suggest that a systematic probing of the role of electrostatics in BLSV droplets should be fruitful.  

\subsubsection{Interfacial} \label{sec:interfacial}

There are at least two generic reasons why interfacial forces may be important. First, large capillary forces operate on the particles in the final stages of the drying of a colloidal suspension when particles poke out of the air-liquid interface, possibly in a liquid film of thickness comparable to the particle diameter.\cite{Kralchevsky1994} These lead to the deformation and coalescence of latex particles,\cite{Sprakel2016} which, in paint, gives rise to a (desirable) continuous film. The stresses involved are of order $\sigma_{\rm dry} \sim \gamma/R$, where $\gamma \lesssim \SI{70}{\milli\newton\per\meter}$ is the surface tension of aqueous solutions while $R \lesssim \SI{100}{\nano\meter}$ is a typical viral dimension, giving $\sigma_{\rm dry} \lesssim \si{\mega\pascal}$. A study has found that similar or larger forces operate when air ingresses into the capsid of bacteriophages in the final stages of desiccation, which may break the capsids and/or eject the genetic material.\cite{Carrasco2009} Two caveats, however, are in order. First, bacteriophages are not enveloped, so this result cannot be applied directly to coronaviruses. Secondly, the numerical modelling in this study neglects electrical forces, which, as we have seen, are non-negligible. 

The second reason why interfacial forces may be important is somewhat less obvious, and arises from flows. Direct viscous forces are likely negligible in viral biophysics. Typical shear rates in, say, a shaking incubator\cite{Giese2014} lie in the range $\dot\gamma \approx 10$ to \SI{e3}{\per\second}. The maximum viscous stress on a virion suspended in an aqueous medium with viscosity $\eta \approx \SI{1}{\milli\pascal\second}$ is therefore of order \SI{1}{\pascal}, which is of the same order as the thermal stresses experienced by a particle of diameter $D \approx \SI{100}{\nano\meter}$, i.e., $\sim k_{\rm B} T/D^3 \approx \SI{1}{\pascal}$, too small to generate any direct mechanical effect.

However, flow may be important indirectly. Intriguingly, it has long been known that when gases are bubbled through a viral solution,\cite{Adams1948} or the solution is tumbled in a test tube\cite{Teppema1974} or flowed through a packed bed,\cite{Thompson1999} viral deactivation ensues. Generically, colloids may partially or completely aggregate under mechanical agitation (stirring, etc.).\cite{Heller1975} However, coagulation {\it per se} does not need to affect viability, although it can decrease the viral titre.\cite{Langlet2007} Moreover, the tumbling inactivation of bacteriophages did not apparently involve coagulation.\cite{Teppema1974} Instead, early experiments show conclusively that the effect is due to viral particles becoming adsorbed at the air-water interface.\cite{Seastone1938} Importantly, since there is in general a significant energy barrier to such adsorption, again possibly due to electrostatics,\cite{Bon2019} adsorption in each of the above-mentioned experimental configurations is likely flow-assisted. 

A 1948 paper suggests that once  a virus is adsorbed at an air-water interface, `it is subjected to such forces that it may very rapidly be deprived of the property of infectivity'.\cite{Adams1948} The nature of these interfacial forces has so far remained obscure, but two well known areas of phenomenology may be relevant. First, there are  interfacial electrical forces~\cite{Kralchevsky2000,Binks2002,Bon2019}, and as we have seen, electrostatics is generically important for viruses. Secondly, and particularly relevant for enveloped viruses, it has long been known\cite{Verger1976,Pattus1978} that bilayer and multilayer lipid vesicles disrupt spontaneously on contact with air-water interfaces and spread out as a monolayer. Importantly, this process still occurs in unilamellar vesicles that include membrane proteins.\cite{Nag1996} The possibility that an enveloped virus may similarly disrupt spontaneously on contact with an air-water interface is intriguing and deserves further investigation, e.g., by studying virions in a Langmuir trough, which has been done for a non-enveloped virus\cite{Nava2016} but not, as far as we know, for enveloped viruses. In this context, as well as more generally, we should also mention that nanoparticles covered by unilamellar lipid bilayers\cite{Vitiello2019} may be useful model systems for aspects of the biophysics of enveloped viruses.

\subsubsection{Osmotic} \label{sec:osmo}

There are also osmotic effects due to the presence of chemical potential gradients, which exert `generalised forces' that drive material fluxes.\cite{Prigogine} How a virus responds to changes to its external osmotic environment therefore depends on the ease with which water and osmolytes can diffuse across its shell. Thus, it has been known for half a century that the response of different non-enveloped viruses incubated at high salt concentration and then subjected to rapid dilution can be explained by the differing permeabilities of their capsids to water and ions. Low permeability to ions leads to a net, rapid influx of water and the bursting of capsids.\cite{anderson1953}. The response of these viruses to such osmotic down shock can be used to infer their mechanical strength.\cite{Cordova2003} 

Turning to enveloped viruses, recall first that water and many other small molecules can diffuse relatively easily across lipid bilayers, which are, however, far less penetrable to ions.\cite{Yang2015} We know of only one relevant in-depth study. Choi et al.~have investigated the effect of osmotic up shock on live as well as whole inactivated influenza viruses,\cite{Choi2015} the latter being the ingredient of some vaccines. Importantly, the effect of chemical inactivation for vaccine preparation are subtle,\cite{Delrue2012} pertaining to modifications of amino and nucleic acids, and not expected to affect significantly the osmotic response.

Choi et al.~report that the internal osmotic pressure of influenza viruses, $\approx 300$~mOsm, is essentially that of physiological saline. When such a virus is suddenly placed in sugar or salt (\ce{NaCl}) solutions of higher osmolarity, static light scattering shows that it shrinks, due to water efflux. In all cases, there is an immediate shrinkage over $\lesssim \SI{5}{\second}$. This is comparable to the time scale for droplet evaporation in air,\cite{Wells1934,Xie2007}  so that virions in air-borne respiratory droplets should show a similar shrinkage response. 

Choi et al. infer the permeability coefficient\footnote{Defined as the flux normalised by concentration gradient and membrane thickness.} of water through the viral membrane from their shrinkage kinetics data; the values obtained, in the range 1 to \SI{6e-4}{\centi\meter\per\second} depending on the osmotic pressure differential, are, interestingly, systematically lower than the value of \SI{3.4e-3}{\centi\meter\per\second} for water permeating through artificial egg phosphatidylcholine (PC) membranes.\cite{Yang2015} Water permeability through lipid membranes increases with the area occupied per lipid irrespective of chain length, saturation or headgroup.\cite{Zeidel2007} Influenza virus membranes have a lower amount of PC relative to phosphatidylethanolamine (PE) than in mammalian cells (average PE to PC ratio $\approx 7$ compared to $\approx 1$ in human cells\cite{Ivanova2015}), and lipids with the PE headgroup [\ce{-PO_3^-CH2CH2N^+H3}] may generically be expected to occupy a smaller area per molecule than those with the PC headgroup [\ce{-PO3^-CH2CH2N^+(CH3)3}].

In the case of hyperosmotic shock using sugar solutions, Choi et al.~observe a second, more gradual, stage of shrinkage after many tens of seconds. The final light scattering signal is erratic and noisy when the osmotic pressure differential is $\gtrsim 500$~mOsm, which the authors take as evidence of membrane destabilisation, correlating well with loss of the ability of inactivated virions to cause immunogenic response. The second stage of shrinkage is absent when viruses are shocked by hypertonic \ce{NaCl} solutions, suggesting ionic leakage through the membrane. The permeability coefficient  of \ce{Na+} through artificial phospholipid membranes is $\gtrsim \SI{e-14}{\centi\meter\per\second}$,\cite{Yang2015} so that any significant leakage must be due to ion channels in the viral envelop, almost certainly self-assembled homo-oligomers of `protein E' in the influenza virus membrane, which is also present in SARS-CoV-2.\cite{Schoeman2019,Arkin2020} Interestingly, protein E ion channels have a marked preference for cations over anions,\cite{Schoeman2019} so that leakage of \ce{NaCl} may be limited by \ce{Cl-} transport.

Intriguingly, Choi et al. find that the addition of carboxymethylcellulose as a viscosity modifier attenuates the osmotic shock response. However, their  explanation that increased viscosity directly lowers the osmotic pressure seems unlikely. 

A number of mysteries therefore remain from this fascinating study. Further work of this kind from a soft-matter perspective should give much relevant insights on the life cycle of such viruses. For example, it would be useful to know the role of osmotic up shock during the drying of BLSV droplets.\footnote{A theoretical suggestion\cite{Minhaz2010} that it is water influx in a {\it hypoosmotic} environment that matters in drying is unlikely to be correct -- the author has neglected to take into account the considerable amount of salt in real respiratory droplets.\cite{Marr2018}} Related to this, Quan et al. has reported that coating the fibres of a mask with salt can inactivate trapped viruses; they attribute this to the effect of osmotic up shock as aerosol droplets come into contact with and dissolve the deposited salt.\cite{Quan2017}  On the other hand, the possible role of osmotic {\it down} shock also deserves investigation. Enveloped viruses are found to be inactivated significantly faster in water than non-enveloped human enteric viruses.\cite{LaRosa2020} This correlates well with the lipid envelope of influenza viruses being an order of magnitude softer than a typical viral protein capsid coat,\cite{Schaap2012} so that they should be more vulnerable to water influx.

\subsubsection{Measurement}

The effect of forces can be studied directly using various kinds of atomic force microscopy (AFM). This has been done for both non-enveloped viruses such as the brome mosaic virus~\cite{Zeng2017} and the cowpea chlorotic mottle virus\cite{Klug2006}, and for enveloped viruses such as the influenza virus~\cite{Schaap2012} and the murine leukemia virus.~\cite{Kol2006} Interestingly, the mechanical properties of vesicles made from influenza virus envelope lipids have also been measured.\cite{Li2011} However, the results may not be directly applicable to understanding real enveloped viruses, because proteins expressed in the lipid bilayer envelope play vital roles in determining membrane mechanics.\cite{Schaap2012} Such `mechanical virology' is a very active area of research\cite{Mateu2012,Roos2020} in which soft matter scientists should be able to make a strong contribution. Progress in  viral mechanics, coupled with an understanding of what forces operate on viruses, especially at interfaces, should suggest strategies by which virions may be mechanically inactivated.

\subsection{Re-entry}

For completeness, we mention that virus-bearing droplets deposited on fomite surfaces must gain entry to and infect a susceptible individual to complete the infection cycle. This is mostly part of the `inside' story, so that we will not discuss the matter in detail. Suffice it to say that the rehydration of a substantially desiccated BLSV composite particle and its wetting kinetics on a mucus-covered epithelial surface will involve complex, and fascinating, soft matter science and biological physics, such as various osmotic effects (Section~\ref{sec:osmo}).

\section{Protection and disinfection}

\subsection{Face masks}

The wearing of face masks for health and safety has a long history. Pliny the Younger (died 79 C.E.) describes how `[p]ersons polishing cinnabar in workshops tie on their face loose masks of bladder-skin, to prevent their inhaling the dust in breathing, which is very pernicious, and nevertheless to allow them to see over the bladders'.\citep{Pliny} The mask made from bladder -- soft matter of living origin -- evidently covered the whole face, but was translucent enough for the wearer to retain adequate vision. 

According to both field surveys\cite{Macintyre2020} and theoretical epidemiological modelling\cite{Stutt2020} published since the start of the current epidemic, the wearing of face masks may offer protection against \mbox{COVID-19} infection. The extent to which this is because the wearing of masks reduces the release of respiratory droplets into the air by infected individuals and/or reduces the inhalation of such droplets by susceptible individuals is not clear. A recent Schlieren imaging study does, however, make it clear that wearing various face coverings can significantly reduce the spatial extent of the frontal air flow ejected by a person while breathing or coughing.\cite{Viola2020,Viola2020b} Different mask designs and materials clearly affect the efficacy of reducing emission,\cite{Verma2020} and the extent of such reduction during speech can be quantified by a new low-cost method.\cite{Fischer2020} We focus our discussion on fabric face masks.

These masks work by filtering out virus-bearing droplets, Fig.~\ref{fig:transmit}(f). In America, the Center for Disease Control (CDC) recommends the use of N95 grade masks for protecting against SARS-CoV-2. N95 filter fabrics are certified according to a protocol set out by the CDC's National Institute for Occupational Safety and Health (NIOSH Document 42 CFR Part 84).\cite{NIOSH,Eninger2008} NIOSH specifies that mask fabrics should be tested for their ability to filter out, at an air flow rate of 85 litres per minute, NaCl aerosols (median diameter $75 \pm \SI{20}{\nano\meter}$, which presumably evaporate rapidly). To qualify as N95 filter fabric, the aerosol concentration downstream must be 5\% of that upstream. The recommendation of N95 grade masks in the current pandemic is presumably based on the fact that the specified median NaCl aerosol size (\SI{75}{\nano\meter}) is somewhat smaller than the virion diameter (\SI{100}{\nano\meter}). (See also related discussion and further references in Bar-On et al.\cite{Milo2020})


An important issue that has emerged as the pandemic crisis progresses is whether face masks could be washed and reused.\cite{Liao2020} The potential soft matter science contribution to this issue is to understand what happens when a BLSV droplet lands on a network of (possibly charged) synthetic polymer fibres and subjected first to humid air flow (wearer inhalation/exhalation) and then to liquid flow and heat (washing), Fig.~\ref{fig:transmit}(g). To arrive at such an understanding, we need progress on practically all of the areas reviewed in Sections~\ref{sec:surfaces} and \ref{sec:forces}, and more. `More' because, now, there is a complex fibrous network to contend with. To take just one of the new issues raised by this complex environment, the study of liquid drops deposited on flexible fibre arrays is only in its infancy.\cite{Duprat2012} The combined action of heat and soaps on natural or synthetic fibre structures also needs to be studied. Note in this context that humidity has an important effect on the permeability of fabric.\cite{Gibson1999} The new physics of why fabric networks actually hold together\cite{Warren2018} may also prove relevant here. The soft matter science of mask cleaning is therefore wide open. 

\subsection{Sanitising}

`Every nurse ought to be careful to wash her hands very frequently during the day. If her face too, so much the better,' so says Florence Nightingale in her pioneering text published in 1859 based on field experience during the Crimean War.\cite{Nightingale1859} Her prescient remarks, made just as Louis Pasteur was beginning his experiments that led to the germ theory of infectious disease,\cite{Berche2012} are as relevant today as they were one and a half centuries ago. A 1999 {\it British Medical Journal} editorial was entitled: `Hand Washing: A modest measure -- with big effects',\cite{Handwashing} while a recent letter to the editor of another journal during the COVID-19 crisis is entitled `Revisiting Nightingale's legacy'.\cite{Dancer2020} The World Health Organisation (WHO) agrees; one of their COVID-19 pamphlets\cite{WHO} is headlined `SAVE LIVES: CLEAN YOUR HANDS', explaining that `COVID-19 virus primarily spreads through droplet and contact transmission. Contact transmission means by touching infected people and/or contaminated objects or surfaces. Thus, your hands can spread virus to other surfaces and/or to your mouth, nose or eyes if you touch them.'

Many aspects of the physical science of hand washing is unknown. In fluid dynamics, for example, the authors of a recent review\cite{Mittal2020} article say, `Amazingly, despite the $170^+$ year history of hand washing in medical hygiene, we were unable to find a single published research article on the flow physics of hand washing.'  The ignorance is less stark when it comes to the soft matter science underpinning hand hygiene: there is a large relevant background literature to form the basis of research specifically targeted at SARS-CoV-2 and other respiratory viruses. 

Hands can be sanitised against such viruses using different chemical agents,\cite{Kampf2020} the most common being surfactants and alcohols. A 2009 study found that, although both are efficacious to varying degrees, washing hands with soap and water was superior against the H1N1 influenza virus.\cite{Grayson2009} Based on results from studies of sanitising hands against bacterial pathogens,\cite{Todd2010} the CDC suggests\cite{CDC} that soap should also be more effective against SARS-CoV-2 in non-clinical settings. The advantage may not, however, be intrinsic. Outside the clinic, hands are more likely to be soiled. Washing with soap is effective in removimg grease and other forms of dirt, which can trap pathogens. 

There are many studies of the action of surfactants on model lipid bilayer systems such as giant unilamellar vesicles (GUVs), some of which are directly inspired by the use of surfactants as agents against enveloped viruses (e.g., in vaginal microbiocides\cite{Vanderlick2003}). A preprint\cite{Chaudhary2020} lists previous publications on the antiviral action of commercial surfactant products, but these and other studies report measurements of efficacy with little to say about mechanism. Significantly, one study\cite{Pauli1997} of the inactivation of enveloped viruses by surfactin, a potent biosurfactant from {\it Bacillus subtilis}, claims that `the antiviral action \ldots seems to be due to a {\it physicochemical interaction} of the membrane-active surfactant with the virus lipid membrane.' The words we have italicised point to soft matter rather than molecular biology.

One interesting example of a mechanistic study is of the inactivation of human (H3N2) and avian (H5N3) influenza viruses by potassium oleate (C18:1), sodium laureth sulfate (LES) and sodium lauryl sulfate (SLS) using isothermal titration (ITC).\cite{Kawahara2018} It found that surfactant-virus interaction was exothermic for LES but endothermic for the other two, with LES being the least effective inactivation agent, and C18:1 being the most effective. The authors speculated on molecular mechanisms, suggesting, for example, that there is strong electrostatic interaction between the negative head group of C18:1 and the positively-charged hemagglutinin proteins embedded in the viral lipid envelope.\cite{Schoeman2019}  Much more work is needed, e.g., paying attention to the aggregation state of the surfactants, to test these suggestions. 

It is convenient to discuss another observation under the heading of surfactants. Unsaturated fatty acids are known to be effective in inactivating enveloped viruses (flu, SARS, \mbox{COVID-19},~\ldots).\cite{Kohn1980} However, the pK$_a$ of the moieties studied are all around 9.5,\cite{Shah2002} so that at or around neutrality, these molecules are fatty acid {\it oils} and not surfactants. Their action must be more subtle, and  not understood at present. 

Surfactants are generally not `kind' to skin; but the long history of using them in personal care products means that there is substantial knowhow in dermatological alleviation.\cite{Walters2012} Another disadvantage of surfactants -- formulated as solid or liquid soaps --  is that they require washing by water. In many situations, such as typically occurring in healthcare, this is either unavailable or at least inconvenient. Here, alcohol-based hand sanitisers come into their own. 

The WHO has a longstanding recommendation for alcohol-based hand sanitisers.\cite{WHO2} The mechanism of alcohol inactivation of enveloped viruses is no better elucidated than is the case for surfactants. Moreover, a recent preprint studying the drying of mixed alcohol-water films finds instabilities\cite{Prash2020} that can lead to the formation of holes in the film and therefore patchy, heterogeneous disinfection. On the other hand, hole formation gives rise to air-liquid interfaces, which can, if coupled with flow, be advantageous for viral inactivation (see Section~\ref{sec:interfacial}). The matter deserves further study.

The WHO formulations, which contain 80\% (v/v) ethanol or 75\% (v/v) isopropyl alcohol and $\lesssim 2\%$ glycerol in water, are low-viscosity Newtonian {\it liquids}. Pouring and rubbing these on hands is inconvenient due to rapid runoff; this presumably underlies the WHO's recommendation to surgeons that `a minimum of three applications are used, if not more, for a period of 3-5 minutes'. 

The solution to this problem is to gel the sanitising liquid. Many polymers are available that will gel aqueous solutions at typically $\lesssim 1\%$ w/v, but  high alcohol content is a challenge, because under these conditions, many if not most of the usual gelling agents are poorly soluble. Some products specifically marketed for solubility at high alcohol concentrations require careful pH adjustment for dissolution, and dissolution is a difficult subject that is far from fully understood.\cite{Koenig2003} Furthermore, at least in the early phase of the COVID-19 crisis, the hand sanitiser industry faced a shortage of the common polymers used in hydro-alcoholic hand gels, exposing the desirability of finding new polymers that can be used as alternatives. There is knowledge on some of the basic soft matter science here, e.g.~in the food literature.\cite{Norton2017} Note further that the effect of these gelling agents on virus inactivation is unknown.

Alcohols also have their own disadvantage: they dissolve lipids from the stratum corneum, the outmost layer of skin. This leads to cracking. Discomfort apart, cracking opens up routes for the entry of (non-respiratory!) pathogens. Repeated use over a long period, necessary for many frontline workers, is therefore a problem. The soft matter science for alleviating this problem has hardly started to develop. Previous work on the microfluidics of skin\cite{Lips2003} and more recent advances in understanding stratum corneum physics\cite{Olmsted2016} and elastocapillarity\cite{Style2017} should form a good basis for advances.

We should mention that nanoparticles are also possible antiviral agents. One recent paper using designer binding ligands on the surface of particles reports the generation of large forces when viruses bind, leading to irreversible deformation.\cite{Cagno2018} This and other uses of nanotechnology-enabled approaches to combating COVID-19 have recently been reviewed.\cite{Weiss2020}. 

Our focus here is on the `outside' story of respiratory viruses. However, one aspect of the `inside' story of such viruses should be mentioned here, because it suggests a novel soft-matter-based strategy for sanitising. It is known that moieties that promote positive membrane curvature inhibit the fusion of viruses with their host cells.\cite{Cheetham1994} This principle has been used to design potential new antiviral drugs.\cite{StVincent} Perhaps such `fusion inhibitors' may also be used to decrease the infectivity of virions on fomite surfaces if a suitable means of application can be devised. 	

So far we have discussed hand sanitising. Many of the issues raised are, however, also relevant to the disinfecting and cleaning of inanimate surfaces. For such surfaces, of course, a broader range of chemical agents may be acceptable than for use on skin, so that novel antiviral mechanisms may operate. The use of ultraviolet (UV) radiation also becomes possible for fomite sanitising. The recommendation\cite{Derraik2020} is to use the UVC band (below \SI{280}{\nano\meter}). Here, it is important to note that both NaCl and proteins absorb strongly below \SI{300}{\nano\meter}, so that viruses encrusted in salt and mucin are likely somewhat protected against UVC.

\section{Detection}

Detecting infection has been at the forefront of public attention since the beginning of the COVID-19 crisis. The topic can conveniently be discussed under the three headings of detecting genetic material, whole viruses, and antibodies/antigens. 

Much of the science of genetic material detection lies in the domain of molecular biology. Soft matter science may, however, still make critical contributions. Thus, soft matter researchers have long studied the microfluidics of complex fluids, which is used, e.g., in making nucleic-acid or antibody based tests available as `point-of-care' personal devices.\cite{Zhang2017} Soft matter physics-inspired approaches can also help improve nucleic acid-based methods for detecting viruses. For example, a recent paper shows that designing oligonucleotide probes to bind multivalently to target bacterial DNA sequences should give better sensitivity and selectivity,\cite{Allen2020} with potential application to RNA and SARS-CoV-2.

Detecting whole viruses can be done in many ways. Imaging by electron microscopy is perhaps the most unambiguous, but relies on perhaps the most expensive equipment. Another way to proceed is by detecting the binding of virions to suitably functionalised colloids by measuring the diffusive dynamics of the latter. Thus, flu viruses can be detected under ideal laboratory conditions through their binding to gold nanoparticles by monitoring the latter's diffusion using dynamic light scattering.\cite{Tripp2011} However, discriminating against false positives when virus hunting in complex body fluids in a clinical setting will be more challenging. Many other novel soft-matter based methods for the efficient detection of whole viruses are possible, e.g., through their binding with pre-stretched DNA bundles embedded in hydrogels and the resulting bulk mechanical deformation.\cite{Metzler2014}

Alteration in particle dynamics can also be used to detect the binding of viral proteins (antigens) or antibodies produced by infected individuals. However, these being small molecules, a highly sensitive method is required. Holographic microscopy has recently been demonstrated as being sensitive enough for this task.\cite{Zagzag2020} One of the inventors of the technique has suggested that it can be used for detecting antibodies from COVID-19 infection or whole virions of SARS-CoV-2.\cite{APS}

For completeness, we mention the state of aggregation of viruses, which must be known, e.g., in order to understand the meaning of viral titre measurements in terms of plague forming units. The state of viral aggregation can be characterised in a variety of ways, some of which are familiar in soft matter science.\cite{Wei2007}

\section{Sustainability}

The COVID-19 pandemic has many direct and indirect environmental consequences.\cite{Saadat2020,Ruano2020}  Some of these impacts are positive, such as dramatically reduced NO$_2$ pollution.\cite{Muhammad2020} The COVID-19 pandemic can be seen as an involuntary experiment to measure the effect of global behavioural change on CO$_2$ emission.\cite{Peters2020} The results show that `social responses alone \ldots would not drive the deep and sustained reductions needed to reach net-zero emissions', but that `structural changes in the economic, transport or energy systems' will be needed.

Other impacts are negative.\cite{Saadat2020,Ruano2020}  Of direct concern to soft matter science is the increase in plastic waste associated with the rise in hand sanitiser and one-off PPE usage. These trends, together with the sudden rise in consumer demand for take-away and individually-packaged food products, mean that the COVID-19 pandemic contributes to the worsening of a parallel `plastic pandemic'. The dramatic increase in the use of hand wipes has also increased non-biodegradable and non-recyclable waste. Soft matter science is a key player in the continued drive to find more biodegradable plastics and other green materials\cite{Miao2017} for all of these applications. On a related note, we have mentioned the potential use of various nanotechnologies in the global response to the COVID-19 (and future) pandemics. As these technologies are being developed, their environmental impact\cite{Wilson2018} must be carefully evaluated, for example, within a `responsible innovation' framework.\cite{Fisher2020,EPSRC}

\section{Discussion and conclusions}

As the authors of a recent review\cite{Mittal2020} have remarked in their conclusion, `[t]he COVID-19 pandemic has exposed significant scientific gaps in our understanding of critical issues, ranging from the transmission pathways of such respiratory diseases, to the strategies to use for mitigating these transmissions.' They have given `a fluid dynamicist's perspective on important aspects of the problem.' In this article, we have given a soft matter scientist's perspective on the knowledge gaps revealed by COVID-19, some of which are closely related to issues discussed in the fluid dynamics essay. 
Given the importance of airborne transmission in the spread of influenza, SARS-CoV-1, Middle East Respiratory Syndrome coronavirus and SARS-CoV-2 (see Zhang et al.\cite{Zhang2020} and references therein), progress in many of the areas we have identified must be seen as urgent, not only for responding to the current crisis, but also to anticipate future respiratory viral pandemics. 

In some cases, progress in the soft matter science of COVID-19 will depend on advances in other, allied, areas. For example, much of soft matter research on mucin to date has used solutions of reconstituted, purified proteins,\cite{Bansil2018} which do not mimic the complex microstructure of `the real thing'; neither do they reproduce {\it in vivo} compositions, e.g., due to the absence of DNA. Until recently, the only solution was to rely on {\it ex vivo} mucus samples obtained from relevant tissues.\cite{Mackie2014} The availability of mucus-secreting organs-on-a-chip \cite{Bhatia2014,Nawroth2020} therefore may offer new opportunities for soft matter work on viruses embedded in realistic mucus droplets and films. 

SARS-CoV-2 and other enveloped respiratory viruses are highly infectious. In the UK's list of approved classification of pathogens, all members of the family Coronavirinae listed are classified into Hazard Group (HG) 3, with the single exception of the human coronavirus 229E, which is in HG~2.  Most soft matter scientists are unlikely to have access to their own HG~2 or 3 facilities.\footnote{At the time of writing, the UK's Advisory Committee on Harmful Pathogens has provisionally classified SARS-CoV-2 into HG 3, so that all laboratory activities must be carried out under Containment Level 3 rules. The CDC's interim guidance is consistent with this, but allows certain procedures (such as diagnostic tests) to be carried out in Containment Level 2 laboratories. These are interim classifications. The reader who needs up to date information should consult the websites of the respective agencies. The complete list of approved classification of various pathogens are available in the UK from the Health and Safety Executive, \url{https://www.hse.gov.uk/pubns/misc208.pdf}, and corresponding agencies in other countries.}   However, in the spirit of much soft matter science applied to biology, it is possible to make progress on many of the research problems we have identified using model systems. The way a low concentration of nanoparticles may affect the fragmentation of viscoelastic liquid jets and sheets\cite{Lindner2015,Raux2020} is a good example -- it is likely that results obtained using \SI{100}{\nano\meter} negatively-charged synthetic colloids should be applicable for understanding infected sneeze ejecta; the more important issue here is likely the non-Newtonian properties of the liquid matrix rather than realistic virions. Another area where model systems work may have high impact is the study of bespoke nanoparticles modified, e.g. by patchiness\cite{Banquy2019} or lipid bilayer coating,\cite{Vitiello2019} to mimic enveloped viruses. Interestingly, nanoliposomes\cite{Mozafari2010} have long been suggested as credible mimics of such viruses,\cite{Martin1974} but we know of little, if any, subsequent work taking this up.

Translating advances in the areas we have described to practical solutions will require collaboration between soft matter scientists and specialists from many disciplines. In particular, work with virologists, epidemiologists and others in the medical community is needed to verify efficacy. On a purely practical level, soft matter scientists who want to work on respiratory coronavirus will probably need to access Containment Level 3 laboratory facilities operated by their microbiology colleagues. Moreover, engagement and collaboration with industry will ensure that solutions can be scaled up to make a significant impact rapidly, particularly for applications to the current pandemic.

Before offering some concluding remarks, we should emphasise that we have not aimed to survey the areas that we have covered with any degree of comprehensiveness -- each area deserves a critical review of its own. There are also areas of soft matter science relevant to the `outside' story of the SARS-CoV-2 virus that we have not been able to touch on; and we have quite deliberately left out most of the `inside' story. But what we have been able to cover should hopefully be enough to convince the reader that there are many immediate opportunities for applying soft matter science to help the global effort in combating the COVID-19 pandemic. 

\setcounter{footnote}{0}

We offer two final remarks to conclude. First, scientific issues thrown up by COVID-19 will not quickly disappear. On the contrary, they are here to stay. We do not know how long SARS-CoV-2 will remain troublesome, but there are now already `second waves' in many places around the world, or even `third waves', and countries in the Northern hemisphere are being urged to prepare for fresh outbreak in the coming winter.\cite{winter} In any case, there is now heightened global awareness of a state of affairs that has in fact existed for some time: that the world is ill prepared for  pandemics in general.\cite{Garrett1995} In particular, the world was and remains ill prepared against corona viruses. The prescient review by Cheng et al.~in 2007\cite{Cheng2007} ended with these words: 
\begin{quote}
The presence of a large reservoir of SARS-CoV-like viruses in horseshoe bats, together with the culture of eating exotic mammals in southern China, is a time bomb. The possibility of the reemergence of SARS and other novel viruses from animals or laboratories and therefore the need for preparedness should not be ignored.
\end{quote}
These words {\it were} largely ignored in the wake of the 2003 SARS outbreak that did not become a pandemic. There is now a global effort to `catch up', not only to defeat \mbox{SARS-CoV-2}, but also to prepare for a next `Disease X'\footnote{This is the WHO's terminology for the next serious international pandemic'.\cite{X}} caused by another coronavirus or some other enveloped virus. Research to fill the scientific gaps we have identified will therefore remain timely for the foreseeable future (and beyond). 	

Secondly, and finally, we note that making progress in solving many of the problems we have highlighted will yield new fundamental soft matter science. This direction of `knowledge transfer', from applications to basics, is not often emphasised in either political rhetoric or academic discourse,\cite{Poon2016} but is becoming increasingly relevant, especially, perhaps, in the `new normal' that will emerge after the pandemic has run its course.

\section*{Conflicts of interest}
There are no conflicts to declare.

\section*{Acknowledgements}
We thank Rosalind Allen, Daan Frenkel and Rebecca Poon for helpful comments on the manuscript, two anonymous referees for pointing out a number of highly-pertinent references and suggesting additional fruitful areas for discussion, and Aidan Poon for designing an earlier version of Fig.~\ref{fig:transmit}. ATB is funded by an EPSRC Innovation Fellowship (EP/S001255/1) and LLN is funded by EPSRC SOFI CDT (EP/L015536/1). ATB, SOLD, CEM, and WCKP are funded by the BBSRC National Biofilm Innovation Centre (BB/R012415/1). 


\balance


\bibliography{COVID} 
\bibliographystyle{rsc} 

\end{document}